\documentclass[conference]{IEEEtran}
\IEEEoverridecommandlockouts
\usepackage{cite}
\usepackage[T1]{fontenc}
\usepackage[utf8]{inputenc}
\usepackage{amsmath,amssymb,amsfonts}
\usepackage{algorithmic}
\usepackage{graphicx}
\usepackage{textcomp}
\usepackage{xcolor}
\def\BibTeX{{\rm B\kern-.05em{\sc i\kern-.025em b}\kern-.08em
    T\kern-.1667em\lower.7ex\hbox{E}\kern-.125emX}}
\begin{document}

\title{ThreatVisionAI: A Hybrid CNN-ViT Framework for Image-Based Malware Classification
}



\author{\IEEEauthorblockN{Allyson Taylor}
\IEEEauthorblockA{\textit{Department of Mathematics \& Computer Science} \\
\textit{University of North Carolina}\\
Pembroke}
\and
\IEEEauthorblockN{Prashanth BusiReddyGari}
\IEEEauthorblockA{\textit{Department of Mathematics \& Computer Science} \\
\textit{University of North Carolina}\\
Pembroke}

}

\maketitle

\begin{abstract}
Traditional malware detection methods struggle to generalize to obfuscated or previously unseen threats. This paper introduces ThreatVisionAI, a hybrid malware family classification framework that integrates a raw-image CNN, a wavelet-based CNN, and a Vision Transformer (ViT) to capture complementary spatial, frequency-domain, and global relational features in malware images. The wavelet-based CNN captures multi-scale frequency information that helps distinguish closely related families, while the ViT branch models long-range dependencies across the image. Evaluated on the Malimg dataset, ThreatVisionAI achieves 98.01\% accuracy and a weighted F1 score of 0.9742, with wavelet-domain features providing measurable gains on minority and visually similar families. These results confirm that frequency-aware and transformer-based representations improve image-based malware family classification. 
\end{abstract}


\section{Introduction}
The cybersecurity threat landscape faces constant pressure from malware that uses obfuscation, polymorphic transformations, and zero-day vulnerabilities to evade traditional defenses \cite{obfus, stego}. Signature-based systems depend on known patterns and fail against unseen threats \cite{sigbased, surveymal}. Heuristic approaches offer broader coverage but produce high false-positive rates that undermine operational trust \cite{sigbased, surveymal}. These limitations motivate solutions that can adapt to modern malware ecosystems while maintaining high accuracy \cite{whole, mlad}.

Image-based malware classification offers a promising alternative. By transforming binary files into visual representations, these methods capture structural relationships useful for distinguishing obfuscated and polymorphic variants at the family level \cite{imcec}. Accurate family classification enables incident responders to reuse knowledge about behavior and countermeasures for related samples, supporting faster triage and more targeted defense.

However, image-based classification faces several challenges. The volume and diversity of malware require scalable solutions that generalize across families and platforms \cite{whole, mlad}. Attackers continuously adapt through obfuscation and camouflage, making reliable family assignment difficult \cite{obfus, stego}. Many AI-driven frameworks also lack interpretability, leaving analysts without clear explanations for model decisions \cite{mltrt, fine-tune}.

Many existing frameworks also fail to exploit frequency information useful 
for distinguishing closely related families. CNNs capture local spatial 
patterns and ViTs model global structure, but both typically discard 
high-frequency details through conventional pooling or strided convolutions 
\cite{selfsup, ftvit}. Recent hybrid CNN--Transformer models such as 
LeViT-MC \cite{LeViTMC} and ConvNeXt--Swin \cite{alshomrani2025} improve 
over single-branch architectures, but every branch in these frameworks 
operates exclusively in pixel space. No existing hybrid integrates a 
dedicated frequency-domain branch alongside both a spatial CNN and a 
global-context Transformer, leaving directional textural differences 
between visually similar families unexploited.

To address these gaps, we propose ThreatVisionAI, a hybrid framework combining a raw-image CNN, a wavelet-based CNN, and a ViT. The wavelet CNN captures spatial and frequency-domain characteristics, the ViT models global dependencies, and weighted soft voting fuses their predictions without additional trainable parameters.

This research makes the following contributions:
\begin{enumerate}
    \item Proposes \textbf{ThreatVisionAI}, a hybrid malware family classification framework combining a raw-image CNN, a wavelet-based CNN, and a ViT through weighted soft voting.
    \item Demonstrates through ablation experiments that wavelet-based features strengthen discrimination between closely related malware families by incorporating frequency-domain information unavailable to spatial filters alone.
    \item Applies Grad-CAM to provide interpretable visual explanations of model behavior, confirming that systematic misclassification of Autorun.K as Yuner.A is driven by inter-class visual similarity rather than model 
    error.
\end{enumerate}

\section{Literature Review}
Traditional malware detection has relied on signature-based methods, which compare files against a database of known patterns to identify threats \cite{mlmd, sigbased}. While effective against known malware, these approaches fail against novel or obfuscated variants \cite{siglimit}. Polymorphic malware alters its code structure upon execution, while metamorphic malware rewrites its entire codebase, both evading fixed-pattern detection \cite{mlmd, surveymal}. Heuristic-based techniques attempt to address this by identifying behaviors indicative of malicious activity, but often produce high false-positive rates, misclassifying benign behavior as malicious \cite{hunt}. Together, these limitations highlight the need for more adaptive detection approaches that can cope with the growing complexity of modern malware threats \cite{obfus}.

Machine learning (ML) has significantly transformed malware detection by providing adaptable and efficient ways to identify complex attack patterns. Unlike signature-based methods, ML-based models can generalize to novel threats, making them particularly effective against polymorphic and previously unseen malware \cite{mlmd}. Supervised learning models, including decision trees and random forests, train on labeled datasets to classify files as benign or malicious based on extracted features \cite{aibasedma}, enabling detection beyond predefined patterns and improving real-time response capabilities \cite{enhancecyberrealtime}. Deep learning architectures such as CNNs have further strengthened malware detection by analyzing binaries as images, identifying intricate and evolving threat characteristics \cite{mlmd}.

As malware grows in complexity, detection methods must integrate multiple advanced techniques to improve accuracy and adaptability. CNNs have been widely used in image-based malware detection, converting binary files into grayscale images to extract spatial patterns indicative of malicious code \cite{maldetect, PalominoImageCNN}. However, CNNs are primarily designed to capture local features and struggle to model long-range dependencies across the malware image \cite{imbalanced, GCViT}. Models such as IMCFN achieve high accuracy in static classification but have limited ability to adapt to rapidly changing threat landscapes \cite{IMCFN}.

ViT-based models address these limitations by using self-attention mechanisms to analyze entire malware images holistically, capturing global dependencies that CNNs miss and making them effective for structurally complex or highly varied malware families \cite{GCViT, LeViTMC, ftvit, combinedspatial}. Nonetheless, ViTs are computationally expensive and typically require large training datasets, making deployment challenging in constrained or near-real-time settings \cite{LeViTMC}.

Hybrid models attempt to combine the strengths of multiple techniques. LeViT-MC integrates CNN components with a lightweight ViT to balance computational efficiency and classification accuracy \cite{LeViTMC}. Recent work has extended this paradigm further: Ashawa et al.\ propose a ResNet-152 architecture for transformer-based malware classification \cite{ashawa2024}, Wang et al.\ introduce MalSort, a lightweight self-supervised Swin Transformer framework achieving 98.28\% accuracy on Malimg \cite{wang2024malsort}, Zhao and Gan augment the Swin Transformer with deformable attention to reach 99.35\% on Malimg \cite{zhao2024malware}, and Alshomrani et al.\ fuse ConvNeXt-Tiny with a Swin Transformer in an explainable hybrid architecture evaluated across Malimg, MaleVis, and VirusMNIST \cite{alshomrani2025}.
These results confirm that combining CNN and Transformer branches consistently outperforms single-branch architectures for image-based malware classification. However, across both earlier models, such as IMCFN \cite{IMCFN} and the most recent hybrid architectures, every branch operates exclusively in pixel space. Spatial filtering and strided convolutions discard high-frequency textural detail that distinguishes closely related families, a limitation directly responsible for confusion between visually similar variants such as Swizzor.gen!E and Swizzor.gen!I \cite{MalwareFreqDomain}. ThreatVisionAI addresses this gap by introducing a wavelet-based CNN as a third branch alongside a spatial CNN and a ViT, capturing frequency-domain structure unavailable to any purely spatial hybrid.

\begin{table}[htbp]
\caption{Representative image-based malware classification models and their architectural characteristics relative to ThreatVisionAI.}
\label{tab:related_models}
\centering
\footnotesize
\begin{tabular}{|l|c|c|c|c|}
\hline
\textbf{Model / Study} & \textbf{CNN} & \textbf{ViT} & \textbf{Freq.-} & \textbf{Hybrid} \\
& & & \textbf{Aware} & \\
\hline
IMCFN \cite{IMCFN} & \checkmark & & & \\ \hline
LeViT-MC \cite{LeViTMC} & \checkmark & \checkmark & & \checkmark \\ \hline
Freq.-domain CNN \cite{MalwareFreqDomain} & \checkmark & & \checkmark & \\ \hline
MalSort \cite{wang2024malsort} & & \checkmark & & \\ \hline
Enhanced ViT \cite{zhao2024malware} & & \checkmark & & \\ \hline
ResNet-152 \cite{ashawa2024} & \checkmark & & & \\ \hline
ConvNeXt + Swin \cite{alshomrani2025} & \checkmark & \checkmark & & \checkmark \\ \hline
\textbf{ThreatVisionAI (ours)} & \checkmark & \checkmark & \checkmark & \checkmark \\ \hline
\end{tabular}
\end{table}

\section{ThreatVisionAI}

ThreatVisionAI classifies malware families from image representations using three parallel branches. The raw-image CNN captures local spatial patterns directly from byte-level visualizations. The wavelet-based CNN extracts multi-scale frequency structure via Haar decomposition, revealing textural cues invisible in raw pixel space. The ViT models long-range dependencies across the full image using self-attention. No single branch covers all three signals simultaneously, which motivates our hybrid design. 

\begin{figure*}[t]
  \centering
  \includegraphics[width=0.866\textwidth]{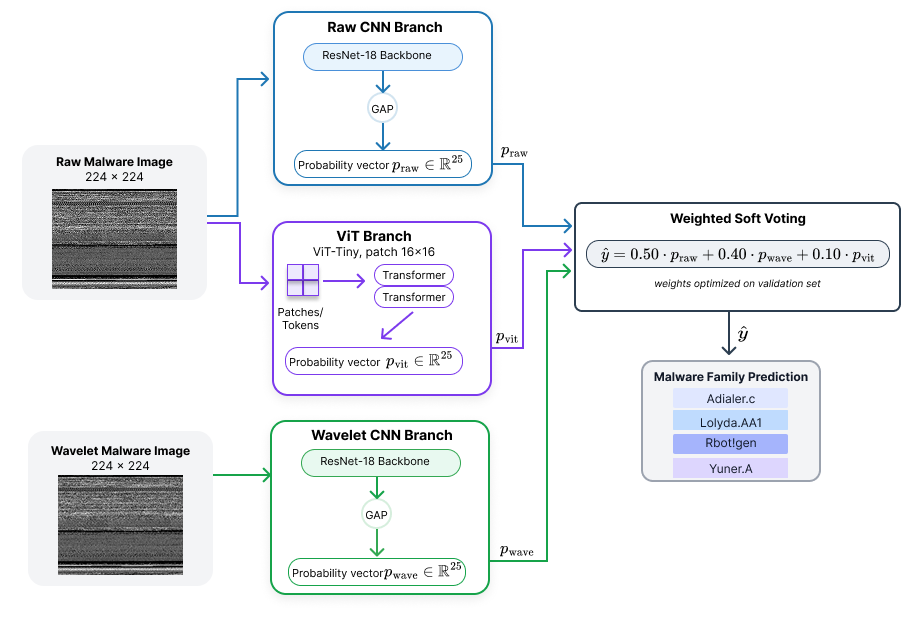}
  \caption{Overview of the ThreatVisionAI architecture.}
  \label{fig:threatvisionai_arch}
\end{figure*}

Each branch is trained independently using the Adam optimizer with a learning rate of $1\times10^{-4}$, class-weighted cross-entropy loss, random rotations up to $\pm15^\circ$, horizontal flips, and brightness jitter of $\pm20\%$, retaining the checkpoint with the highest validation weighted F1. Each branch produces a 25-class probability vector over the Malimg families, combined via weighted soft voting with weights of 0.50 (raw CNN), 0.40 (wavelet CNN), and 0.10 (ViT) selected by exhaustive grid search. Figure~\ref{fig:threatvisionai_arch} shows the full architecture.

\subsection{Convolutional Neural Network (CNN)}

The raw-image CNN branch processes grayscale malware images directly, learning spatial features from the byte-level visual structure of each binary. Malware binaries converted to images exhibit distinctive visual textures, repeated byte patterns, and structural regularities that vary across families \cite{imcec, malimg}. A CNN trained on these images learns these patterns hierarchically, from low-level edges and textures in early layers to family-specific structural signatures in deeper layers.

This branch uses a ResNet-18 backbone \cite{resnet}, a well-established 18-layer deep residual network that introduces skip connections between layers to mitigate the vanishing gradient problem and enable stable training of deeper architectures. ResNet-18 was chosen for its strong representational capacity, computational efficiency relative to deeper variants, and proven effectiveness on image classification tasks at this scale. The network is trained from scratch with no pretrained weights, since ImageNet-pretrained weights are optimized for natural photographs and do not transfer cleanly to the structured, grayscale byte patterns in malware images. The input convolutional layer is adapted to accept single-channel $224\times224$ grayscale images. The final fully connected layer is replaced with a 25-way output head corresponding to the 25 Malimg malware families. Global average pooling over the final feature maps produces a 512-dimensional representation before the classification head.

\subsection{Wavelet-Based CNN}

The wavelet-based CNN branch processes malware images in the frequency domain, exposing textural and structural cues that are not visible in raw pixel space and that help separate families sharing similar global appearance. Each $224\times224$ grayscale image is decomposed using a single-level two-dimensional Haar wavelet transform, which applies low-pass and high-pass filters along both spatial dimensions to produce four coefficient sub-bands: the approximation sub-band (cA), which retains low-frequency content and resembles a downsampled version of the original image, and three detail sub-bands capturing high-frequency variation in the horizontal (cH), vertical (cV), and diagonal (cD) directions. Each sub-band is half the spatial resolution of the input, yielding four $112\times112$ coefficient maps. These four maps are stacked as channels to form a single four-channel $112\times112$ input tensor, giving the model simultaneous access to coarse structure and fine directional detail within a single forward pass.

This branch uses the same ResNet-18 backbone as the raw-image CNN \cite{resnet}, trained from scratch under identical conditions. The input convolutional layer is modified to accept four-channel inputs, with all weights randomly initialized and trained from scratch. The final fully connected layer is replaced with a 25-way classification head, and global average pooling produces a 512-dimensional feature vector consistent with the raw-image branch.

\subsection{Vision Transformer}

The Vision Transformer branch models long-range spatial dependencies across the malware image, capturing relationships between distant regions that convolutional filters cannot reach regardless of depth. CNNs process images through local receptive fields that grow incrementally across layers, so relating two distant regions requires propagating information through many intermediate layers, which is indirect and lossy. ViTs address this by dividing the input into a fixed grid of non-overlapping patches, projecting each patch into an embedding, and applying multi-head self-attention across all patches simultaneously \cite{ftvit}. Every patch attends directly to every other patch in a single layer, making global context immediately available throughout the network.

This branch uses ViT-Tiny pretrained on ImageNet-21k via the \texttt{timm} library. ViT-Tiny was selected for its computational efficiency relative to larger ViT variants, which is important given the limited size of the Malimg dataset and the risk of overfitting a high-capacity transformer on fewer than 10,000 samples. Pretraining on ImageNet-21k provides general-purpose patch-level representations that transfer more reliably to malware images than training from scratch, where the data volume is insufficient to learn stable attention patterns.

The ViT branch takes the same four-channel $112\times112$ wavelet-transformed inputs as the wavelet CNN (cA, cH, cV, cD). Standard ViT architectures expect three-channel inputs, so the patch projection layer is modified to accept four channels, with the additional channel weights initialized by averaging the pretrained three-channel weights and fine-tuning from there. The [CLS] token output from the final transformer encoder block serves as the global image representation and feeds a 25-way classification head.

\subsection{Hybrid Model}

The hybrid model combines the three branches through weighted soft voting rather than learned feature-level fusion. Each branch is trained independently to convergence and produces a 25-class probability vector via its softmax output head. At inference, the three probability vectors are combined as a weighted sum to produce the final class scores, with the predicted family selected as the argmax of the combined vector.

Weighted soft voting was chosen over feature concatenation and MLP-based fusion because it requires no additional trainable parameters, keeps each branch's contribution fully transparent, and avoids the risk of overfitting a fusion head on a small dataset. Branch weights of 0.50 (raw CNN), 0.40 (wavelet CNN), and 0.10 (ViT) were selected by exhaustive grid search over the validation weighted F1.

\section{Experiments and Results}

Experiments are conducted on the Malimg dataset \cite{malimg}, a widely used benchmark for image-based malware classification in which malware binaries are converted to grayscale images by mapping each byte to a pixel intensity value. The dataset contains 9,465 images spanning 25 malware families, with class sizes 
ranging from 80 to 2,949 samples, reflecting a significant imbalance. The largest family, Allaple.A, contains 2,949 samples, while several 
families such as Autorun.K, Skintrim.N, and Yuner.A contain fewer than 100 samples each. The data is partitioned 70/15/15 into training (7,562 samples), validation (946 samples), and test (957 samples) sets, stratified by family to preserve class distribution across all splits. No samples are shared across splits. All models are trained on the training set, hyperparameters, and checkpoints are selected on the validation set, and final performance is reported exclusively on the held-out test set.

Table~\ref{tab:ablation} shows test set accuracy and weighted F1 for each branch in isolation and for two ensemble configurations. The wavelet CNN outperforms the raw CNN on both metrics, with weighted F1 improving from 0.9662 to 0.9733. This gain reflects the value of frequency-domain features for distinguishing families that share similar spatial structure in raw pixel space. Although the ViT branch underperforms in isolation, it contributes complementary global context not captured by CNNs, as reflected in the consistent performance gain when included in the hybrid model \cite{ftvit, taka}. To quantify the ViT's contribution to the full hybrid, we evaluate a two-branch ensemble using only the raw CNN and wavelet CNN branches with renormalized weights. The two-branch ensemble achieves a weighted F1 of 0.9733, matching the wavelet CNN standalone. Adding the ViT branch advances weighted F1 to 0.9742, confirming a small but consistent and repeatable gain from incorporating the global self-attention representation that neither CNN branch captures independently. A CNN and ViT ensemble without the wavelet branch achieves a weighted F1 of 0.9692, below the CNN + Wavelet ensemble (0.9733), confirming that 
frequency-domain features contribute more to family discrimination than 
global self-attention alone at this dataset scale.

\begin{table}[t]
\caption{Ablation results on the Malimg test set.}
\label{tab:ablation}
\centering
\footnotesize
\begin{tabular}{|l|c|c|}
\hline
\textbf{Model} & \textbf{Accuracy} & \textbf{Weighted F1} \\
\hline
Raw CNN (ResNet-18) & 0.9729 & 0.9662 \\ \hline
Wavelet CNN (ResNet-18) & 0.9791 & 0.9733 \\ \hline
ViT-Tiny & 0.9572 & 0.9492 \\ \hline
CNN + ViT (no Wavelet) & 0.9760 & 0.9692 \\ \hline
CNN + Wavelet (no ViT) & 0.9791 & 0.9733 \\ \hline
\textbf{Full Hybrid (0.50/0.40/0.10)} & \textbf{0.9801} & \textbf{0.9742} \\ \hline
\end{tabular}
\end{table}

The wavelet branch produces the most pronounced per-class improvements on the Swizzor family variants. Swizzor.gen!E F1 improves from 0.7368 to 0.8571 and Swizzor.gen!I improves from 0.5263 to 0.8276 when comparing the raw CNN to the wavelet CNN. These two families share strong visual similarity in raw pixel 
space, making them difficult to separate using spatial filters alone. The Haar wavelet decomposition exposes directional frequency differences between them that are not visible in the raw image, enabling substantially more reliable discrimination between closely related variants.

Across all three branches, the majority of the 25 Malimg families achieve near-perfect classification, with errors concentrated on a small number of families. The raw CNN correctly classifies all test samples for 20 of 25 families. Errors are confined to Autorun.K (12 samples, all misclassified as Yuner.A), the two Swizzor variants (which exhibit mutual confusion), and minor spillover from C2LOP.P. This pattern confirms that the primary classification challenges in Malimg arise from inter-family visual similarity rather than from model capacity limitations on minority classes.

\begin{figure}[t]
  \centering
  \includegraphics[width=\columnwidth]{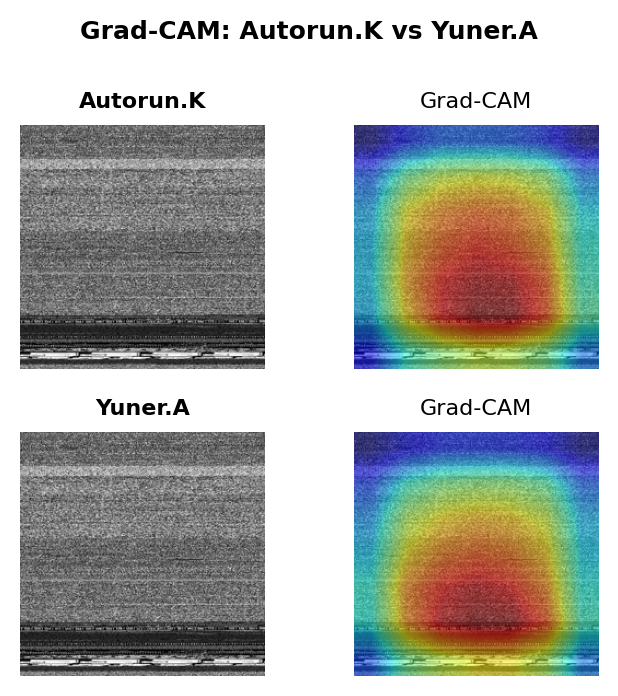}
  \caption{Raw malware images and Grad-CAM heatmaps for Autorun.K (top) and 
  Yuner.A (bottom). The near-identical visual structure and activation patterns 
  confirm that the systematic misclassification of Autorun.K as Yuner.A is 
  driven by inter-class visual similarity rather than model error.}
  \label{fig:gradcam}
\end{figure}

Grad-CAM \cite{gradcam} is applied to the raw CNN branch following recent work on hybrid CNN-Transformer malware classifiers \cite{alshomrani2025}. Figure~\ref{fig:gradcam} shows heatmaps for one Autorun.K and one Yuner.A sample. Autorun.K is the single consistent failure case across all configurations, with F1 remaining 0.0000 in every setting — all 12 test samples are misclassified as Yuner.A. The raw binary visualizations of the two families are nearly identical, and the Grad-CAM activation patterns are indistinguishable, confirming the model attends to the same regions regardless of the true label. Autorun.K and Yuner.A are both Worm-type malware whose visualizations are indistinguishable even to human experts \cite{gibert, dbfsmc}, making this a fundamental property of the dataset rather than a modeling deficiency.

To assess robustness against adversarial evasion, ThreatVisionAI is 
evaluated under the Fast Gradient Sign Method (FGSM) \cite{fgsm} at 
perturbation magnitudes $\varepsilon \in \{0.01, 0.03, 0.05\}$. FGSM 
computes the gradient of the loss with respect to the input image and 
perturbs each pixel by $\varepsilon$ in the direction that maximizes 
classification error, representing a white-box evasion attempt in which 
the attacker has full knowledge of the model. Table~\ref{tab:fgsm} reports 
hybrid accuracy under each perturbation level. Accuracy degrades from 
98.01\% under clean inputs to 59.46\% at $\varepsilon = 0.01$ and 7.84\% 
at $\varepsilon = 0.05$, confirming that ThreatVisionAI is vulnerable to 
white-box adversarial perturbation in its current form. This result 
motivates adversarial training as a direct extension of this work.

\begin{table}[t]
\caption{ThreatVisionAI hybrid accuracy under FGSM white-box evasion.}
\label{tab:fgsm}
\centering
\footnotesize
\begin{tabular}{|c|c|}
\hline
\textbf{Perturbation ($\varepsilon$)} & \textbf{Hybrid Accuracy} \\
\hline
0.00 (clean) & 0.9801 \\ \hline
0.01         & 0.5946 \\ \hline
0.03         & 0.1118 \\ \hline
0.05         & 0.0784 \\ \hline
\end{tabular}
\end{table}

Table~\ref{tab:efficiency} reports parameter count and per-image inference time for each branch and the full hybrid, measured using CUDA event timing. Each ResNet-18 branch contains 11.19M parameters, and ViT-Tiny contributes 5.53M, yielding 27.91M total across the ensemble. Sequential inference requires 5.4\,ms per image on GPU, which is operationally acceptable for malware scanning workflows not subject to hard real-time constraints.

\begin{table}[t]
\caption{Parameter count and inference time per branch and full hybrid.}
\label{tab:efficiency}
\centering
\footnotesize
\begin{tabular}{|l|c|c|}
\hline
\textbf{Branch} & \textbf{Params (M)} & \textbf{ms / image} \\
\hline
Raw CNN (ResNet-18)      & 11.19 & 1.3 \\ \hline
Wavelet CNN (ResNet-18)  & 11.19 & 1.3 \\ \hline
ViT-Tiny                 &  5.53 & 2.8 \\ \hline
Full Hybrid (sequential) & 27.91 & 5.4 \\ \hline
\end{tabular}
\end{table}

\section{Conclusion}

Most existing image-based malware classification frameworks rely exclusively on spatial features, leaving frequency-domain structure and global relational cues unexploited. ThreatVisionAI addresses this by combining a raw-image CNN, a wavelet-based CNN, and a Vision Transformer through weighted soft voting. The wavelet branch exposes directional textural differences invisible to spatial filters, the ViT branch relates distant image regions via self-attention, and soft voting fuses these signals without additional trainable parameters.

The ablation study demonstrates that each design decision is justified by measurable performance gains. The wavelet CNN outperforms the raw CNN by 0.0071 weighted F1, with the largest improvements on visually similar families where spatial features alone are insufficient. Adding the ViT to the two-branch ensemble produces a further consistent gain, confirming that global self-attention contributes a complementary signal that neither CNN branch captures independently.

Grad-CAM analysis provides visual evidence that the single persistent failure case across all configurations, Autorun.K, is not a modeling deficiency but a fundamental property of the dataset. The binary visualizations of Autorun.K and Yuner.A are near-identical, and the activation patterns the model produces for both families are indistinguishable. No image-based classifier can resolve this ambiguity without signal beyond the raw pixel representation. This analysis shows that ThreatVisionAI produces not only accurate but also interpretable predictions, giving analysts visual evidence to understand and validate model decisions.

ThreatVisionAI has several limitations that inform future work. Evaluation is conducted on Malimg alone, a dataset collected in 2011 whose 9,465 samples do not reflect the scale or diversity of contemporary malware ecosystems. Generalization to larger and more recent benchmarks such as MaleVis and BIG2015 remains unvalidated. Adversarial robustness analysis under FGSM confirms vulnerability to white-box evasion, and no testing against stronger attacks, such as PGD or dataset shift, has been conducted. Future work should address these gaps through multi-dataset evaluation, adversarial training against PGD-based perturbations, exploration of alternative wavelet bases and multi-level decompositions, adaptive ensemble weighting, and integration of dynamic behavioral signals toward a more comprehensive multi-modal detection framework.

\bibliographystyle{IEEEtran}      
\bibliography{references}        

\end{document}